\documentclass[prl, amsfonts, twocolumn, nofootinbib, showpacs]{revtex4}
\usepackage{graphicx, epsfig}

\newcommand{\be}{\begin{equation}}
\newcommand{\ee}{\end{equation}}
\newcommand{\bea}{\begin{eqnarray}}
\newcommand{\eea}{\end{eqnarray}}

\newcommand{\gapp}{\mathrel{\raise.3ex\hbox{$>$}\mkern-14mu \lower0.6ex\hbox{$\sim$}}}
\newcommand{\lapp}{\mathrel{\raise.3ex\hbox{$<$}\mkern-14mu \lower0.6ex\hbox{$\sim$}}}
\def\bbox{{\,\lower0.9pt\vbox{\hrule \hbox{\vrule height 0.2 cm
\hskip 0.2 cm \vrule height 0.2 cm}\hrule}\,}}

\begin{document}
\title{Gravitational Lenses in Generalized Einstein-Aether theory: the Bullet Cluster}
\author{De-Chang Dai$^1$, Reijiro Matsuo$^2$, and Glenn Starkman$^2$}
\affiliation{$^1$HEPCOS, Department of Physics, SUNY at Buffalo, Buffalo, NY~~14260-1500\\ $^2$CERCA, Department
of Physics, Case Western Reserve University, Cleveland, OH~~44106-7079}


\begin{abstract}

\widetext
We study the lensing properties of an asymmetric mass distribution and vector field in
Generalized Einstein-Aether (GEA) theory. As vector field fluctuations are responsible
in GEA for seeding baryonic structure formation,
vector field concentrations can exist independently of baryonic matter.
Such concentrations would not be expected to be tied to baryonic matter except gravitationally,
and so, like dark matter halos, would become separated from baryonic matter in interacting
systems such as the Bullet Cluster.
These vector field concentrations cause metric deviations that affect weak lensing.
Therefore, the distribution of weak lensing deviates from that which would be
inferred from the luminous mass distribution, in a way that numerical calculations
demonstrate can be consistent with observations.
This suggests that MOND-inspired theories can reproduce weak lensing observations,
but makes clear the price: the existence of a coherent large-scale fluctuation of a field(s)
weakly tied to the baryonic matter, not completely dissimilar to a dark matter halo.

\end{abstract}


\maketitle

\section{Introduction}

It is well known that General Relativity can not explain
the full dynamics of our universe and the structures in it
sourced solely by visible mass.
Two kinds of exotic energy density, dark matter and dark energy,
must be introduced to explain respectively the dynamics of
structures and the accelerating expansion of the universe.

\begin{figure*}[t]
\centering
\begin{minipage}[t]{8.5 cm}
\begin{center}
\includegraphics[width=3.2in]{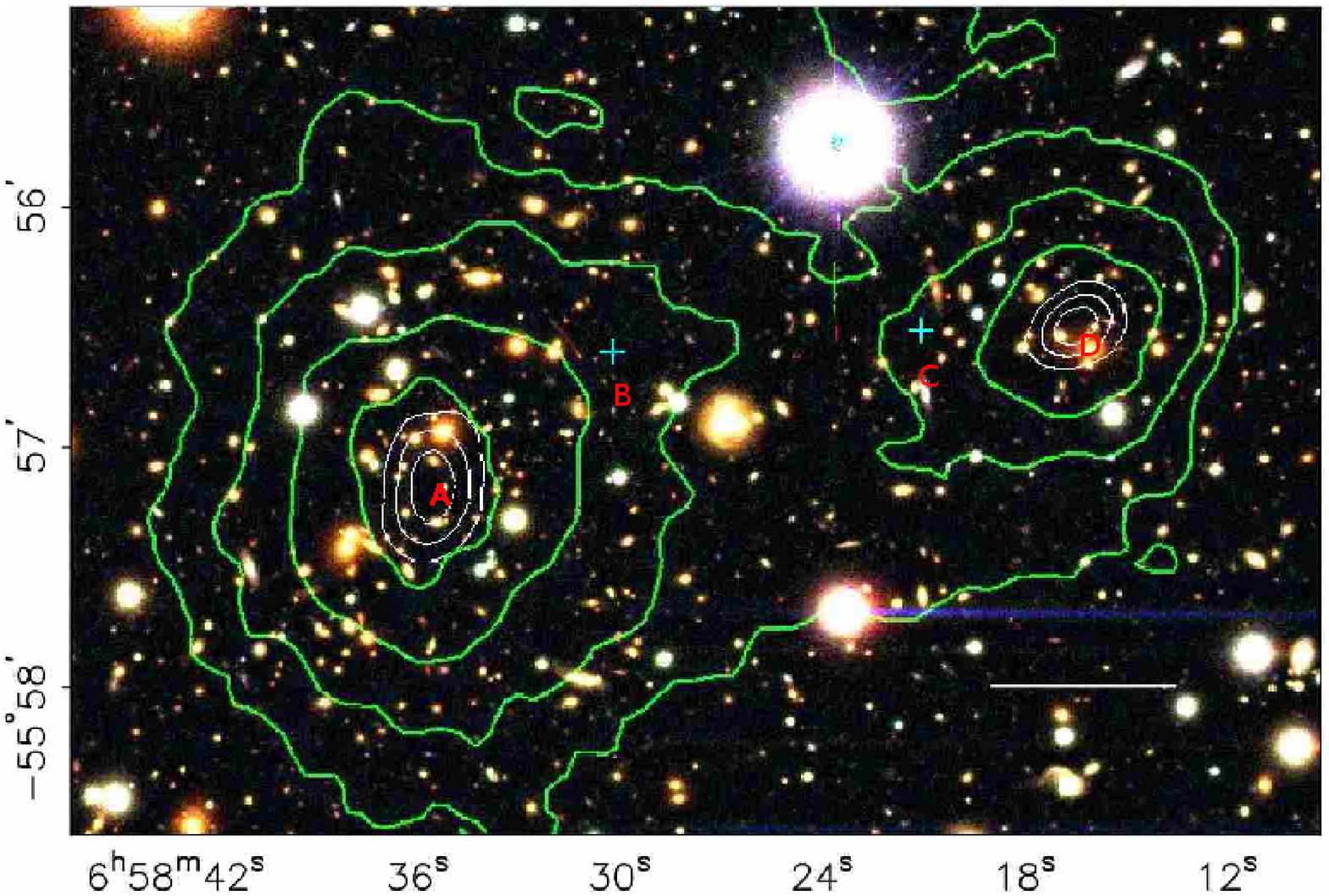}
\end{center}
\end{minipage}
\hfill
\begin{minipage}[t]{8.5cm}
\begin{center}
\includegraphics[width=3.2in]{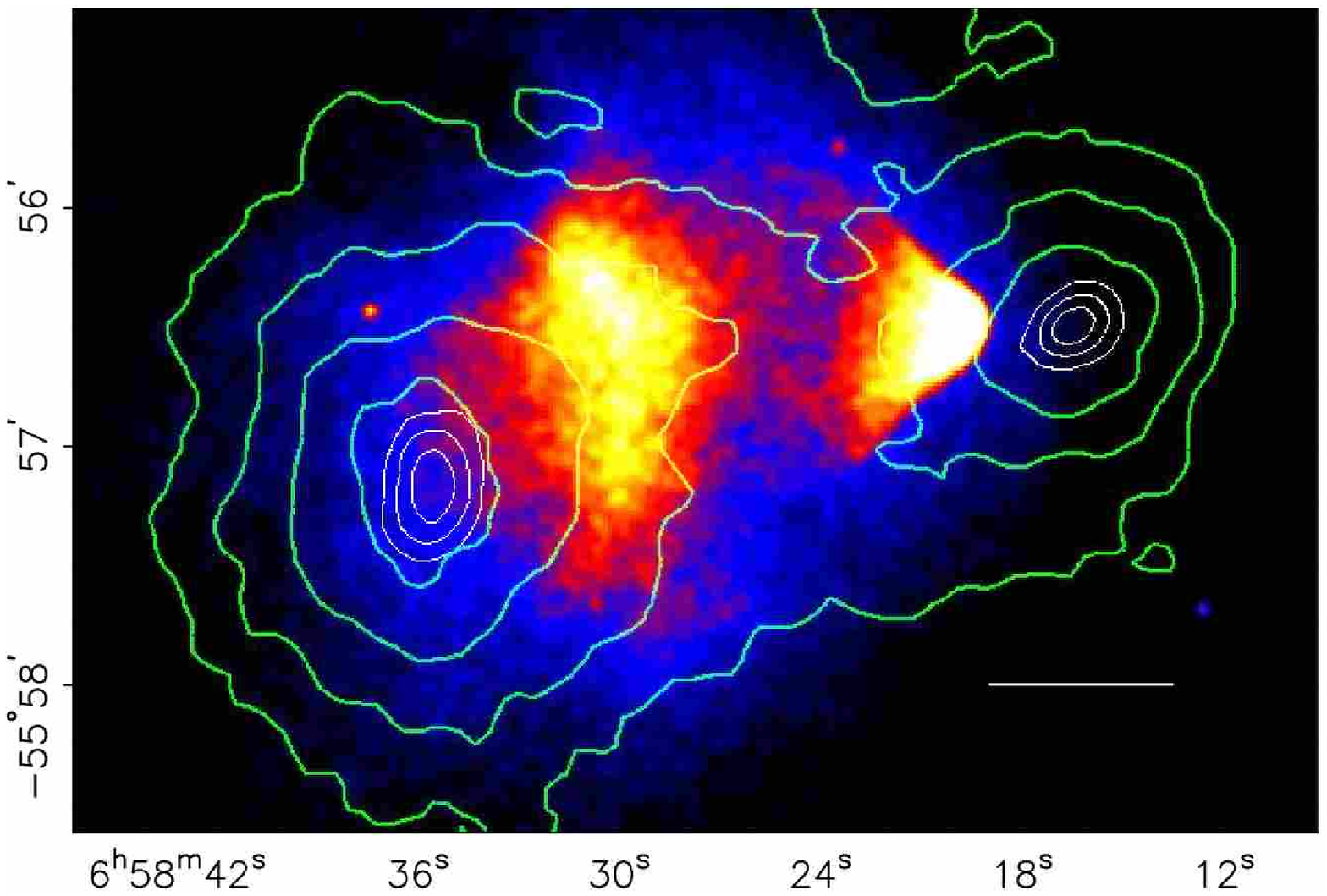}
\end{center}
\end{minipage}

\caption{The bullet cluster\cite{bullet}:
Shown above in the left panel is a color image from the Magellan images
of the merging cluster 1E0657$-$558, with the white bar indicating
200 kpc at the distance of the cluster. In the right panel is a 500 ks
Chandra image of the cluster. Shown in green contours in both panels
are the weak lensing $\kappa$ reconstruction with the outer contour
level at $\kappa =$ 0.16 and increasing in steps of 0.07. The white
contours show the errors on the positions of the $\kappa$ peaks and correspond
to $68.3\%$, $95.5\%$, and $99.7\%$ confidence levels. }
\label{fig:bulletcluster}
\end{figure*}


An alternative to adding new sources of gravity is to alter the
response of gravity to the known matter sources. Modified theories
of gravity replacing dark matter as an explanation of flat galaxy
rotation curves were first proposed by Milgrom
\cite{Milgrom:1983ca,Milgrom:1983pn} and by Milgrom and
Bekenstein \cite{Milgrom:1983ca,Bekenstein:1984tv}, and
came to be known as Modified Newtonian Dynamics a.k.a. MOND
(for a review of MOND see \cite{Sanders:2002pf}).
In the contemporary incarnation of this phenomenological theory,
Poisson's equation for the Newtonian potential $\phi$, is altered
by the inclusion of a scalar function $\mu$:
\be
\nabla\cdot\left(\mu\left(\vert\nabla\phi\vert\right)\nabla\phi\right) = 4\pi G_N\rho .
\ee
The scalar function is chosen to enhance gravitational acceleration at large distance,
and thus replace galactic dark matter. It turns out that galactic dynamics
be explained using very simple choices of $\mu$, such as
\be
\label{eq:muofx}
\mu\left(\vert\nabla\phi\vert\right) \equiv \frac{\vert\nabla\phi\vert/a_0}{\sqrt{1 + \vert\nabla\phi\vert^2/a_0^2}}.
\ee
Introducing only one new parameter, a universal acceleration, $a_{0}$,
one can fit almost all galactic rotation curves assuming very reasonable mass-to-light
ratios consistent with prior expectations.

Despite its successes, most scientists did not regard MOND as a viable theory.
In part this was due to difficulties explaining clusters and possibly some
small scale structures, but more particularly it was due to the absence
of a consistent covariant action formulation of the theory.
More recently, these theoretical hurdles were overcome by Bekenstein \cite{Bekenstein:2004ne},
who proposed Tensor-Vector-Scalar gravity (TeVeS). Zlosnik {\it et al.}\cite{Zlosnik:2006sb}
then showed that TeVeS is part of a larger class that they call Generalized Einstein-Aether (GEA) theories.
In these theories, a new field or fields modify the response of the metric to the presence of matter.
Most generic is the existence of a vector field with a non-zero time-like background value.
Hence the name Einstein-Aether.

It has been claimed that the discovery of the merging galaxy cluster 1E0657-56 \cite{Markevitch:2001ri},
the bullet cluster, rules out these modified gravity theories and proves the existence of dark matter \cite{bullet}.
X-ray images of this cluster (left hand side of figure \ref{fig:bulletcluster})
reveal a bullet-like subcluster just exiting the core of the main cluster.
Weak lensing mass contours (right hand side of Figure \ref{fig:bulletcluster})
show that the lensing centers are not at the luminous centers of these bright clusters,
rather they are (approximately) on the line through these X-ray centers but outside them (cf. Figure \ref{fig:bulletcluster}).
The dark matter interpretation is that the lensing centers (which are less luminous
than the X-ray centers) are caused by the presence of collisionless cold dark matter (CDM).
In the aftermath of the cluster-cluster collision,
the CDM halos have presumably continued along un-impeded,
while the hot baryonic gas has been slowed.

It is true that MOND is unlikely to explain the separation of the gravity centers from the mass centers.
But MOND was never to be regarded as a complete theory, and certainly not as a theory of time-dependent mass distributions.
Previous authors \cite{Angus:2006ev,Feix:2007zm} have taken the MOND-derived Newtonian potential,
inserted it into TeVeS to show that TeVeS too does not explain the observed weak lensing map
without an additional dark mass component.
This has led the Bullet cluster to be widely regarded as direct evidence for dark matter
and against modified gravity.

But are relativistic versions of MOND truly ruled out by the bullet cluster?
The new dynamical degrees of freedom of these theories are known to solve other clear problems of vanilla-MOND.
For example, it was shown \cite{Lue:2003ky} that perturbations glows too slowly in
a MOND universe for the observed large scale structures to have emerged from the
fluctuations present at the time of recombination. However, it has been shown in both
TeVeS \cite{Skordis:2005xk, Dodelson:2006zt} and GEA 
growing modes of the new dynamical fields introduced in these modified gravity theories
successfully seed the growth of baryonic structures after the end of silk damping.
This implies that observed baryonic structures -- galaxies, clusters, {\it etc.} --
evolved from non-baryonic seeds which were not sourced by baryons.
If these seeds persist, then they might be expected to behave in very much the same way as
a dark matter halo when the baryonic structures with which they are associated collide --
travelling on, in a straight line unimpeded. These non-baryonic structures may
then be the non-luminous weak lensing centers observed in the bullet cluster.
Of course, this would mean that even in modified gravity theories baryonic structures
are surrounded by non-baryonic halos -- but halos of weakly interacting classical fields
rather than halos of weakly interacting non-relativistic particles.

For our calculations, we will focus on the Generalized Einstein-Aether theory,
a generalized form of "Einstern-Aether" theory \cite{Eling:2004dk,Jacobson:2000xp}.
In this theory, general relativity is modified only by the addition of a
dynamical vector field $A^\alpha$, with a time-like classical value $A^\alpha=(1,0,0,0)$.
We modify the conventional ansatz for $A^\alpha$ in the presence of a matter source
to allow for first-order perturbations in the spacelike, rather than just the time-like components.
The spacelike component is not sourced by the matter, but could emerge from the
intrinsic vector growing mode.
The vector field enhances weak lensing by different factors at different locations,
with a resulting possible dissociation of the luminous and gravitational-lensing centers.
MOG is another possible modified gravity theory htat might accomodate the
observations of the bullet cluster \cite{Brownstein:2007sr}.

We organize the paper as follows. We first review briefly the Generalized Einstein-Aether theory
and calculate the solution for the vector field to first order. We then discuss appropriate boundary
conditions for the bullet cluster, and perform suitable numerical calculations of the
resulting metric and convergence.

\section{\bf Theory}
\indent
The action of the Generalized Einstein-Aether theory can be written in the form \cite{GEA}
\begin{eqnarray}
S=\int d^4x \sqrt{-g}\left[\frac{R}{16\pi G_N}+{\cal L}(g,A)\right]
+S_{M} . \label{genaction}
\end{eqnarray}
\noindent
Here, \textbf{g} is the usual metric (with signature (-,+,+,+)),
$R$ the Ricci scalar of that metric,
and $S_M$ the matter action.
\textbf{A} is a new dynamical vector field.
As shown in \cite{GEA,Zlosnik:2006sb},
we can obtain a MOND-type limit in the appropriate low-acceleration regime
by choosing
\be
\label{eq:Lagrangian}
{\cal L}(A,g)=\frac{M^2}{16\pi G_N}
{\cal F}({\cal K}) +\frac{1}{16\pi G_N}\lambda(A^\alpha A_\alpha+1),
\ee
where
\begin{eqnarray}
\label{eq:Kdefn}
{\cal K}&=&M^{-2}{\cal
K}^{\alpha\beta}_{\phantom{\alpha\beta}\gamma\sigma}
\nabla_\alpha A^{\gamma}\nabla_\beta A^{\sigma} \quad {\rm and}\nonumber \\
{\cal
K}^{\alpha\beta}_{\phantom{\alpha\beta}\gamma\delta}&=&c_1g^{\alpha\beta}g_{\gamma\sigma}
+c_2\delta^\alpha_\gamma\delta^\beta_\sigma+
c_3\delta^\alpha_\sigma\delta^\beta_\gamma
\end{eqnarray}
The $c_i$ are dimensionless constants, whereas $M$ has the dimension of mass.
$\lambda$ is a non-dynamical
Lagrange-multiplier field with dimensions of mass-squared,
that enforces that $A^\alpha$ is unit-timelike.
The gravitational field equations obtained by varying $g^{\alpha\beta}$ \cite{CL,HEHL}) are
\begin{equation}
G_{\alpha\beta}=\tilde{T}_{\alpha\beta}+8\pi G_NT^{matter}_{\alpha\beta}
\label{fieldI}
\end{equation}
where the stress-energy tensor for the vector field is given by
\begin{eqnarray}
\tilde{T}_{\alpha\beta} &=& \frac{1}{2}\nabla_{\sigma}
({\cal F}'(J_{(\alpha}^{\phantom{\alpha}\sigma}A_{\beta)}-
J^{\sigma}_{\phantom{\sigma}(\alpha}A_{\beta)}-J_{(\alpha\beta)}A^{\sigma}))\nonumber \\
&& -{\cal F}'Y_{(\alpha\beta)}
+\frac{1}{2}g_{\alpha\beta}M^{2}{\cal F}+\lambda A_{\alpha}A_{\beta}
\end{eqnarray}
\noindent where
\begin{eqnarray}
{\cal F}' &\equiv& \frac{d{\cal F}}{d{\cal K}} \quad {\rm and}\nonumber \\
J^{\alpha}_{\phantom{\alpha}\sigma} &\equiv&
(\cal{K}^{\alpha\beta}_{\phantom{\alpha\beta}\sigma\gamma}+
\cal{K}^{\beta\alpha}_{\phantom{\beta\alpha}\gamma\sigma})\nabla_{\beta}A^{\gamma} .
\end{eqnarray}
Brackets around indices denote symmetrization. $Y_{\alpha\beta}$ is the functional derivative
\begin{eqnarray}
Y_{\alpha\beta} &=&
\nabla_{\sigma}A^{\eta}\nabla_{\gamma}A^{\xi}
\frac{\delta(\cal{K}^{\sigma\gamma}_{\phantom{\sigma\gamma}\eta\xi})}{\delta
g^{\alpha\beta}} \\
&=&-c_{1}\left[ (\nabla_{\nu}A_{\alpha})(\nabla^{\nu}A_{\beta})-(\nabla_{\alpha}A_{\nu})(\nabla_{\beta}A^{\nu})\right]
\nonumber
\end{eqnarray}

The equations obtained by varying $A^{\beta}$ are
\be
\nabla_{\alpha}({\cal F}'J^{\alpha}_{\phantom{\alpha}\beta})+{\cal F}'y_{\beta}=2\lambda A_{\beta} ,
\ee
\noindent with
\be
y_{\beta}\equiv\nabla_{\sigma}A^{\eta}\nabla_{\gamma}A^{\xi}\frac{\delta(\cal{K}^{\sigma\gamma}_{\phantom{\sigma\gamma}\eta\xi})}{\delta A^\beta} .
\ee
For the choice of ${\cal K}$ given by equation \ref{eq:Kdefn}, $y_\beta=0$.

Variations of $\lambda$ fix
\be
\label{eq:unittimelike}
A^\mu A_\mu=-1 .
\ee

The MOND type solutions of the Generalized Einstein-Aether theory had been studied in \cite{GEA}.
Only the $c_1$ term in ${\cal K}^{\alpha\beta}_{\gamma\delta}$ plays an important role.
To simplify the problem we choose $c_{2}=c_{3}=0$.
Since we are studying the weak gravity regime
and the velocities of galaxies are much smaller than the speed of light,
we assume that the time derivatives of the gravitational and vector fields are much smaller than their space derivatives.

We expand the vector field and geometric metric around a fixed, Minkowski space background:
\begin{eqnarray}
g_{\alpha \beta}=\eta_{\alpha \beta} + h_{\alpha \beta}, \nonumber \\
A^{\alpha}=\delta ^{\alpha}_{0}+ B^{\alpha},\nonumber
\end{eqnarray}
where $h_{\alpha\beta}$ and $B^{\alpha}$ are taken to be small.
We consider only terms that are linear in $h$ and $B$,
and choose Poisson gauge
\be
h_{00}=-2\Phi \quad\quad h_{ij}=-2\delta_{ij}\Psi .
\ee
$\Phi$ and $\Psi$ are scalar potentials.

The first order term in equation (\ref{eq:unittimelike}) fixes
\begin{eqnarray}
B^{0}&=&-\Phi ,\nonumber\\
{\rm as\quad well\quad as}&&\nonumber\\
\bigtriangledown_{t}A^{t}&=&0\nonumber\\
\bigtriangledown_{t}A^{i}&=&\partial_{i}\Phi \\
\bigtriangledown_{i}A^{t}&=&0\nonumber\\
\bigtriangledown_{j}A^{i}&=&\partial _{j}B^{i}\nonumber\\
{\cal K}&=&M^{-2}c_{1}(-(\partial_{i} \Phi )^2+(\partial_{i}B^{j})^2)\nonumber
\end{eqnarray}
We look for a solution such that $\partial_{i}B^{j}=\partial_{j}B^{i}$,
so that the first order contributions to the off-diagonal elements of the stress-energy tensor,
$\tilde{T}_{ti}$ and $\tilde{T}_{ij}$, disappear.
The $ij$th component of the Einstein equations ($i\neq j$) then yields
\be
\Phi=\Psi .
\ee
This is consistent with General relativity.
The $00$th term of the Einstein equations is
\begin{equation}
\bigtriangledown^{i}[(1+\frac{c_{1}}{2}{\cal F'(K)})\bigtriangledown_{i}\Phi ]=4\pi G_N \rho .
\end{equation}
This motivates us to identify
\be
\mu \equiv 1+\frac{c_{1}}{2}{\cal F'(K) } .
\ee
If $\partial_{i}B^{j}=0$, then we recover the MOND form of gravity
at small accelerations (i.e. at large distances).
Here, we construct a solution that approaches the MOND behavior at small acceleration
even though $\partial_{i}B^{j}\neq0$.
We choose the functional form of $\mu$
(really valid only for the acceleration due to a point mass)
as in (\ref{eq:muofx})
\be
\label{eqn:mu}
\mu = \frac{x}{\sqrt{1+x^2}} ;
\ee
however, now
\be
\label{eqn:xGEA}
x^2=\frac{((\bigtriangledown\Phi)^2-(\partial_{i}B^{j})^2)}{a_{0}^2}\frac{1}{1-p} .
\ee
In principle, $p$ is an arbitrary function of {\cal K},
but we take it to be a constant.
Note that
\begin{equation}
(\bigtriangledown \Phi)^2\geq(\partial_{i}B^{j})^2,
\end{equation}
so the presence of an inhomogeneous vector field establishes a minimum
acceleration.
Moreover, the closer $p$ is to 1, the greater the influence of the vector field.
The usual MOND-type solution is recovered if
\begin{equation}
\label{db}
(\partial_{i}B^{j})^2=p(\bigtriangledown \Phi)^2
\end{equation}
We will take this condition as a starting point.
Ultimately, one must demonstrate that the vector-field dynamics
leads to this relation, at least approximately.


\section{Method}

Outside a single point mass the gravity in the deep MOND regime is
\begin{equation}
\label{eqn:MONDianPhi}
\bigtriangledown \Phi = -\frac{\sqrt{G_Nma_0}}{r}\hat{r}
\end{equation}
The solution of the vector field for equation (\ref{db}) is then
\begin{equation}
\label{Bvec}
\vec{B}=\pm \sqrt{\frac{pG_Nma_0}{2}}\hat{r}
\end{equation}
Equation (\ref{Bvec}) is not well defined at $r=0$;
however, since we are not going to discuss the acceleration near the center,
we ignore this divergence.

Of more concern is that the vector field $A^\alpha$ should approach $(1,0,0,0)$,
in other words $\vec{B}\to0$, at spatial infinity.
It seems that this requires that either the gravity
deviates from the MOND limit (equation \ref{eqn:MONDianPhi}) at very large $r$,
or $p(r)\to0$ at large $r$. At any rate, we shall assume that
MOND applies on cluster scales, and that $p$ remains constant
on such scales, so that any deviations from these two approximations
occurs on scales much larger than the size of a cluster.

As has been shown in \cite{GEA_glow},
perturbations of the vector field grow with the cosmological scale factor.
Structure is formed as the vector field perturbations
attract matter particles and thereby cause the gravitational potential to grow.
The size and the mass of a galaxy or a cluster depends on the size and amplitude of
the vector field perturbation.
The fate of the seed vector perturbations, however, is unstudied.
If these spatial terms of the vector field do not decay after the cluster structure is formed,
then the space surrounding the cluster should still have a vector field that is different
than the value that would be inferred by solving the vector field equation with the
matter as a source. We assume that the vector field perturbation extends over a region that
is larger than the size of our simulation, about $8Mpc$.

For two point particles with mass $m_1$ and $m_2$,
the gravitational acceleration in the MOND regime superposes as:
\begin{equation}
(\bigtriangledown \Phi)^2 \sim (\bigtriangledown \Phi_1)^2+(\bigtriangledown \Phi_2)^2=\frac{G_Nm_1 a_0}{r^2}+\frac{G_Nm_2 a_0}{r^2} .
\end{equation}
Note that this is very different than the usual Newtonian superposition
of $\bigtriangledown\Phi$.
To satisfy this relation we assume that the vector fields adding according to:
\[\begin{array}{l}
(\partial_{i}B^{j})^2= (\partial_i B_1^j)^2 + (\partial_i B_2^j)^2 ,
\end{array}
\]
where $B_1$ and $B_2$ are the vector fields associated with (but not necessarily
sourced by) $m_1$ and $m_2$.

There are several numerical methods \cite{MOND, MOND_code1, MOND_code}
that can be used to calculate the gravitational acceleration in the
environments of a mass distribution according to MOND.
To simplify this problem, we choose an axially symmetric mass distribution.
The method in \cite{MOND} is more suitable and easier for this case.
The code solves the following equations:
\begin{eqnarray}
\bigtriangledown \cdot \vec{U}=4\pi G_N \rho \\
\bigtriangledown \times \nu \vec{U} =0 .
\end{eqnarray}
Here
\begin{eqnarray}
\vec{U}\equiv\mu \bigtriangledown \Phi\\
\bigtriangledown \Phi\equiv\nu \vec{U}
\end{eqnarray}
serve as definitions of $\vec{U}$ and $\nu$ respecitively.
Here, $\mu$ is defined by equations (\ref{eqn:mu}) and (\ref{eqn:xGEA}) .
On a cylindrical grid, defined by $\{(r_i,z_j)\}$ (where $i$ indexes
the radial coordinate, $r$, and $j$ indexes the axial coordinate $z$,
and the symmetry allows us to ignore the azimuthal coordinate),
these differential equations can be replaced by
the following difference equations:
\begin{eqnarray}
\label{eqn:NE}
(U^z_{i,j+1}-U^z_{i,j})\pi (r^2_{i+1}-r^2_i)& &\\
+2\pi \delta z_j(r_{i+1}U^r_{i+1,j}-r_iU^r_{i,j})&=&4\pi m_{ij}\nonumber\\
\label{eqn:curl}
\eta _1 U^r_{i,j}-\eta _2 U^r_{i,j-1}+\eta _3 U^z_{i-1,j}-\eta _4 U^z_{i,j}&=&0 .
\end{eqnarray}
Here $U^r_{i,j}$ is the $r$ component of $\vec{U}$ at $r_i$, $z_j$;
similarly $U^z_{i,j}$ is the $z$ component.
The $\eta_k$ are discrete representations of $\mu dl$,
{\it eg.} $\eta_1 =(\nu_{i-1}\delta r_{i-1}+\nu_{i,j}\delta r_i)/2$.
For details of the calculational method, please refer to \cite{MOND}.

Since our goal is to investigate whether the weak lensing maps
of the bullet cluster can be reproduced in the MOND-like limit of GEA,
we must calculate those maps. The lensing convergence is \cite{lensing}
\begin{equation}
\kappa=\frac{D_{ds}D_d}{c^2 D_s}\int \nabla^2 \Phi d \ell
\end{equation}
Here, $D_{d}$, $D_{ds}$ and $D_{s}$ are the angular distances between
the observer and the lens, the lens and the source, and the observer and the source, respectively.
$\ell$ is the distance along the light ray from the observer.
Once we have solved equations (\ref{eqn:NE}) and (\ref{eqn:curl}),
we can straightforwardly calculate $\kappa$.

\section{Model and Result}

\begin{figure*}[t]
\centering
\begin{minipage}[t]{8.5 cm}
\begin{center}
\includegraphics[width=3.2in]{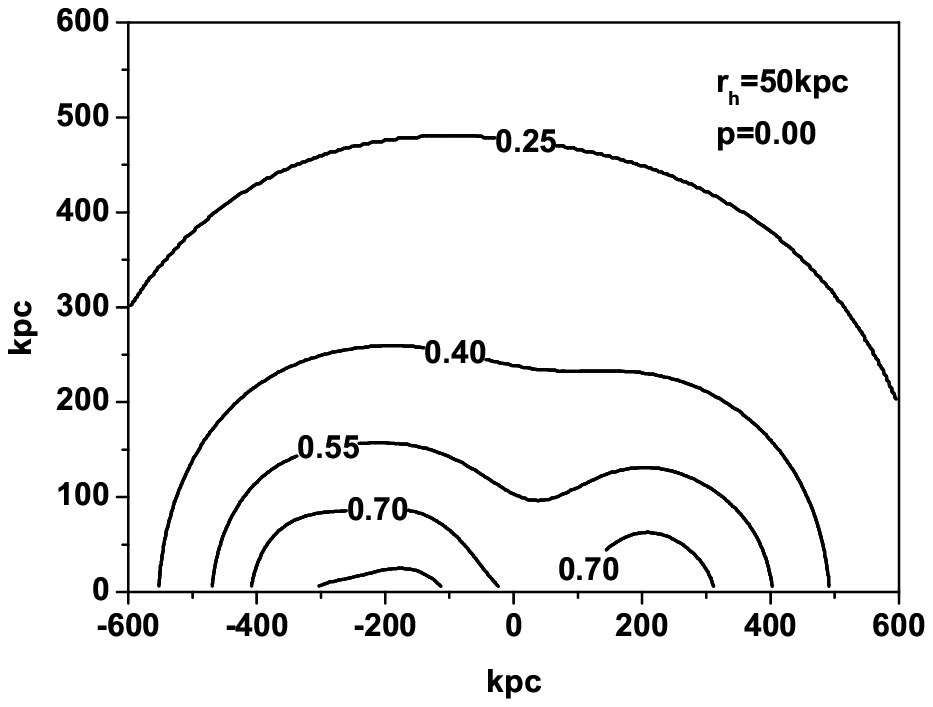}
\end{center}
\end{minipage}
\hfill
\begin{minipage}[t]{8.5cm}
\begin{center}
\includegraphics[width=3.2in]{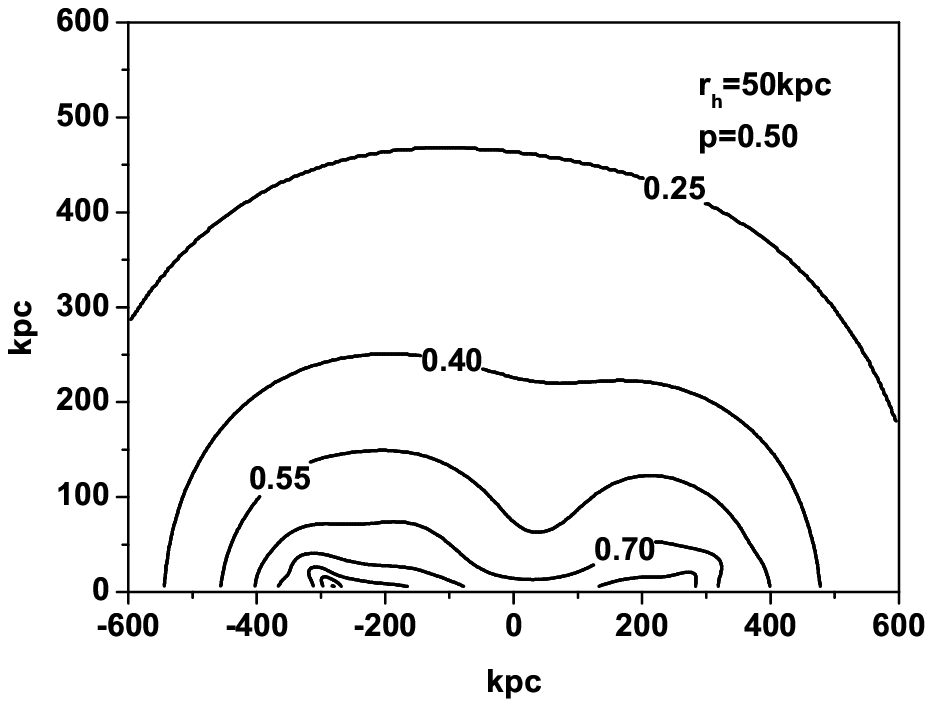}
\end{center}
\end{minipage}
\\
\begin{minipage}[t]{8.5 cm}
\begin{center}
\includegraphics[width=3.2in]{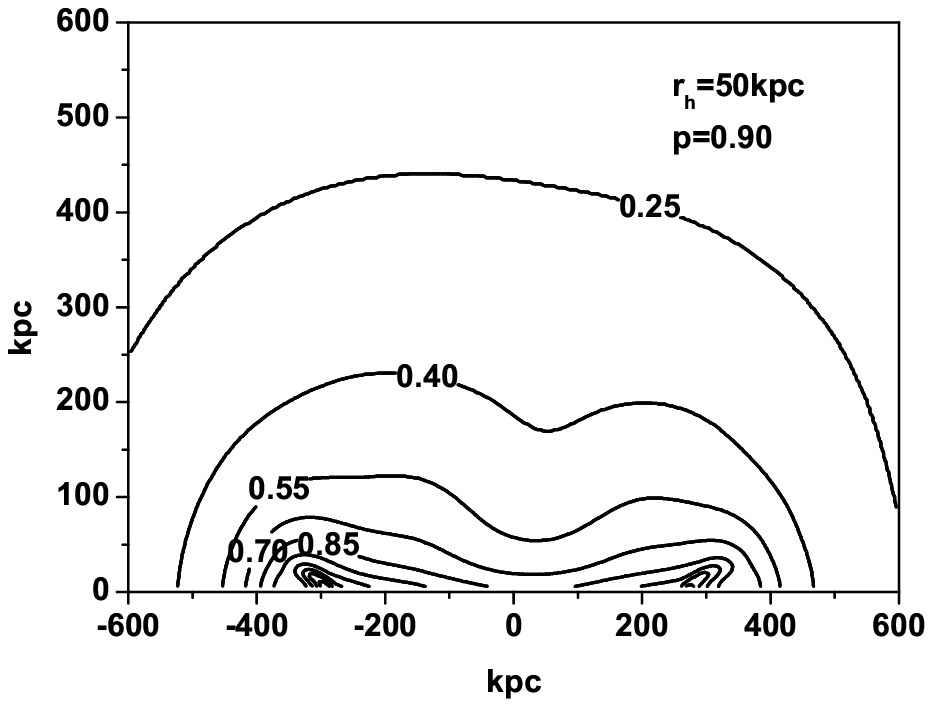}
\end{center}
\end{minipage}
\hfill
\begin{minipage}[t]{8.5cm}
\begin{center}
\includegraphics[width=3.2in]{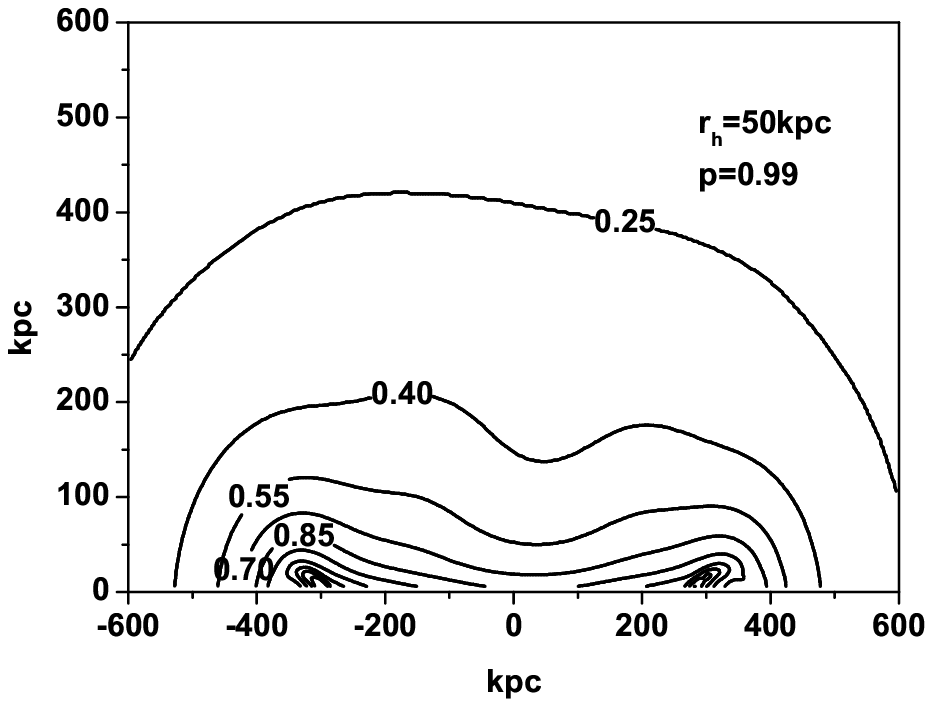}
\end{center}
\end{minipage}

\caption{The covergence map changing with different p: $r_{h}=50kpc$ for the vector fields.
$p=0.00$, $p=0.50$, $p=0.9$, and $p=0.99$ are corresponding to left up, right up, left bottom, and right bottom. The interval between each contour is 0.15. Left up figure shows the result of MOND. It does not show a gravitational lensing center at $-350kpc$ and $350kpc$. As the p increases,
concentracted contours appears.}
\label{change-p}
\end{figure*}

\begin{figure*}[t]
\centering
\begin{minipage}[t]{8.5 cm}
\begin{center}
\includegraphics[width=3.2in]{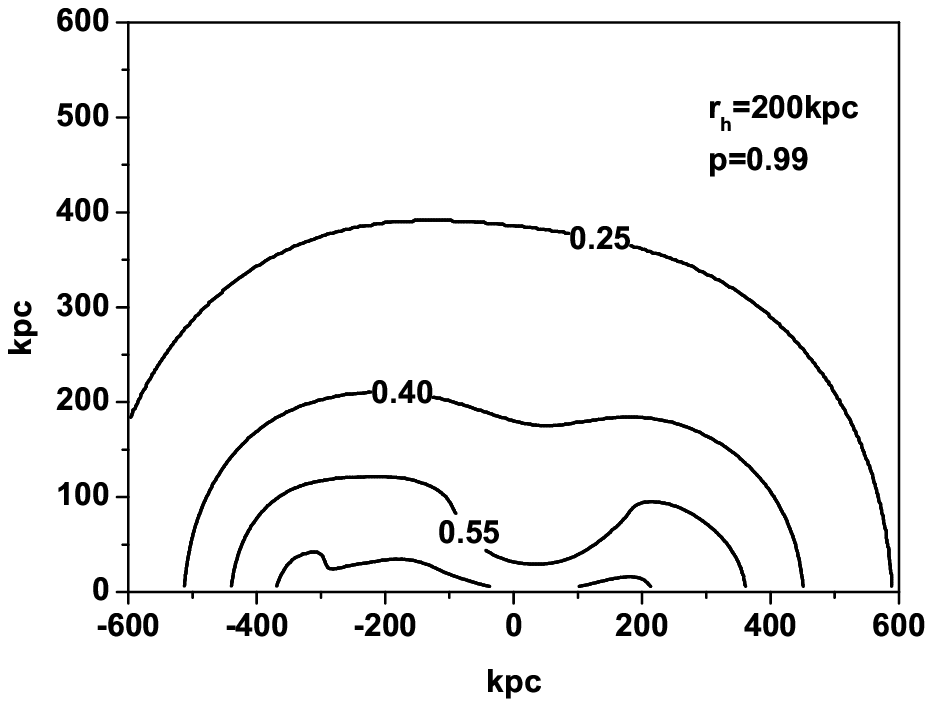}
\end{center}
\end{minipage}
\hfill
\begin{minipage}[t]{8.5cm}
\begin{center}
\includegraphics[width=3.2in]{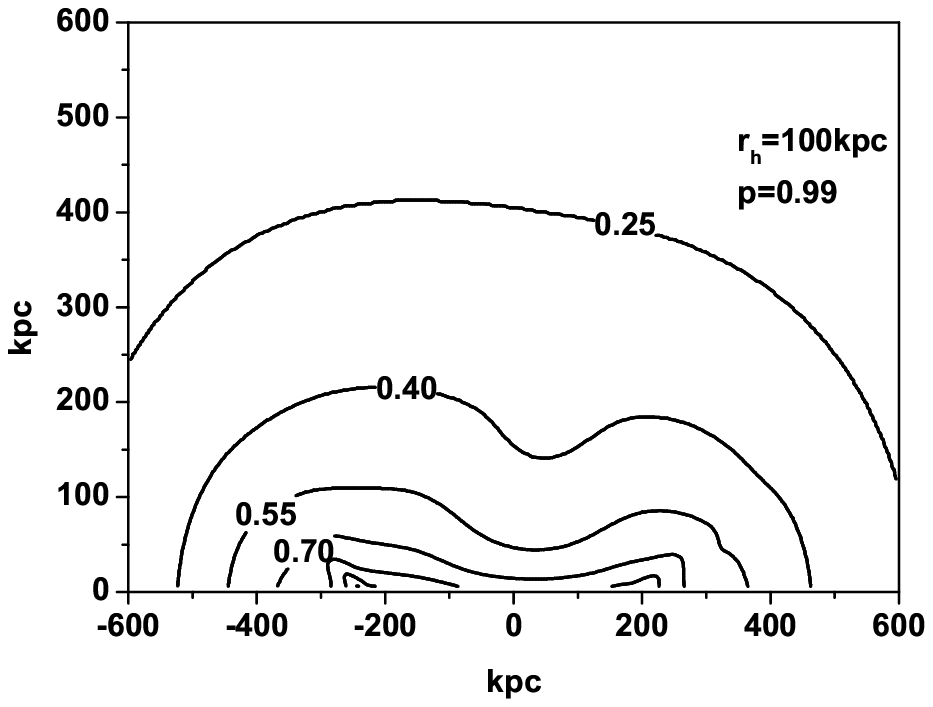}
\end{center}
\end{minipage}
\\
\begin{minipage}[t]{8.5 cm}
\begin{center}
\includegraphics[width=3.2in]{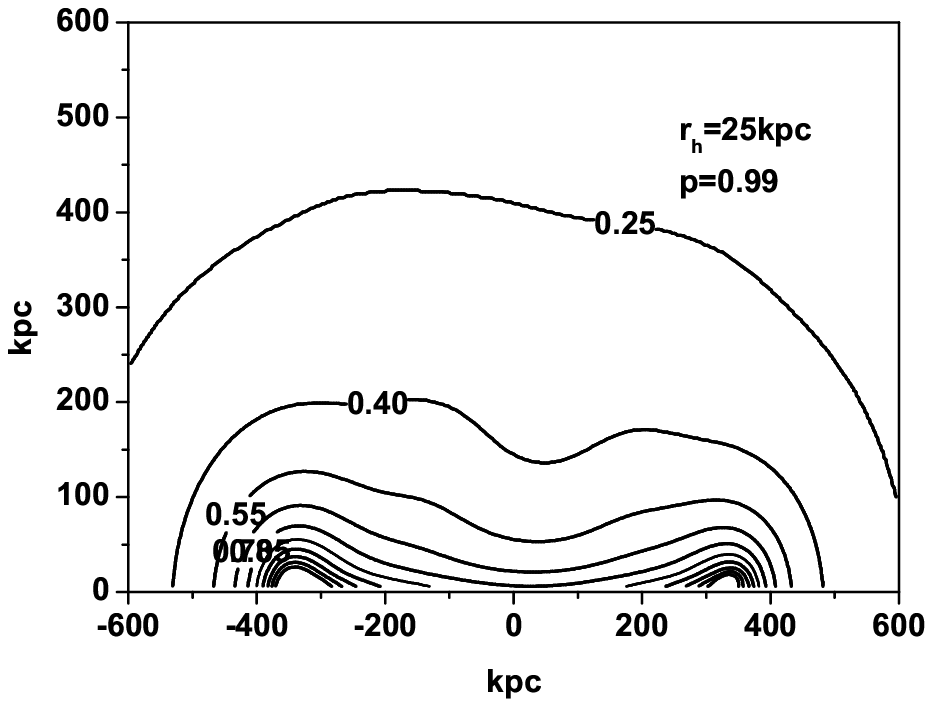}
\end{center}
\end{minipage}
\hfill
\begin{minipage}[t]{8.5cm}
\begin{center}
\includegraphics[width=3.2in]{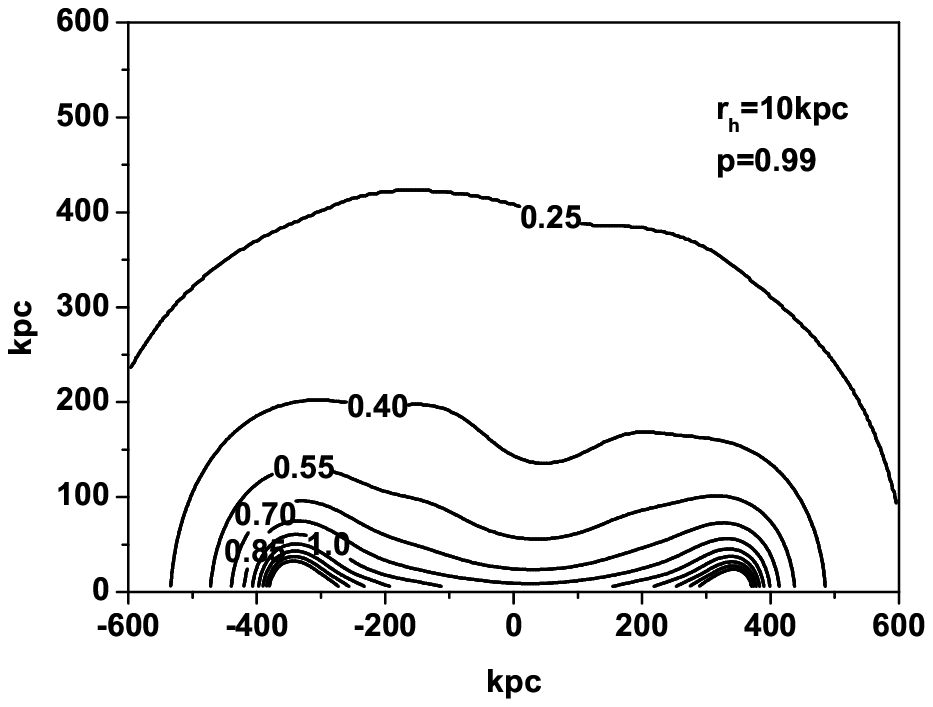}
\end{center}
\end{minipage}

\caption{The covergence map changing with different $r_h$ for the vector fields: $p=0.99$ for the model.
$r_h=200kpc$, $r_h=100kpc$, $r_h=25kpc$, and $r_h=10kpc$ are corresponding to left up, right up, left bottom, and right bottom.
The interval between each contour is 0.15. As the $r_h$ decreases, concentracted contours appears near $-350kpc$ and $350kpc$.}
\label{change-rh}
\end{figure*}

Our code is able to simulate only axially symmetric mass distributions.
The mass distribution of the bullet cluster, fortunately, is very close
to axially symmetric. The angle between the direction of the motion of the
bullet cluster and the the plane of the sky is about $8^o$.
We, therefore, assume that the bullet cluster is axially symmetric
about an axis that is parallel to the plane of the sky.

For the baryonic mass distribution of the cluster, we take a King profile \cite{king}:
\begin{equation}
\label{king}
\rho(r)=\frac{\rho_0}{\left[1+(\frac{r}{r_c})^2\right]^{\frac{3}{2}}} .
\end{equation}
We choose a core radius, $r_c=150 kpc$.
The central density, $\rho_0$, and the positions of mass centers
are chosen acording to table \ref{mass-distribution} as extracted from
\cite{bullet}.
Figure \ref{fig:bulletcluster}
We choose $1.5Mpc$ to be the cutoff of the mass distribution.
$\frac{D_{ds}D_{d}}{D_s}=540Mpc$, which is appropriate for the lensed galaxies
in the background of the bullet cluster, where
the source of the gravitational lensing is at $z_l=0.296$,
and the a mean redshift of the lensed galaxies is $z_{bg}=0.85$

In addition to the baryonic clusters, we add vector field configurations at position A and D.
The only coupling constants in the model are the usual gravitational coupling $G_N$,
and the very small acceleration $a_0$. We therefore assume that the
interaction between the vector field ``halos'' associated with each cluster are sufficiently small
that the two vector-halos can pass through each other unimpeded, much as dark matter halos are assumed to.
The mass term in equation (\ref{eqn:MONDianPhi}) for the vector field at A is
chosen to be the mass of component A plus that of component B.
Similarly, the mass term in equation (\ref{eqn:MONDianPhi}) for the vector field at D is
chosen to be the mass of component C plus that of component D.
We assume the mass also follows equation (\ref{king}).
But the $r_h$s of the vector fields are different from those of the
baryonic mass distributions.

\begin{table}[h]
\caption{The mass and position of baryon matter}
\label{mass-distribution}
{\small
\begin{tabular}{|c|c|c|c|}
\hline
component & type& position & Plasma mass \\
& & (kpc) & within $100kpc$\\
\hline
A & lensing & -350 & 6.0 $\times 10^{12}M_{\odot}$\\
& center & & \\
\hline
B & X-ray & -150 & 6.8 $\times 10^{12}M_{\odot}$\\
& center & & \\
\hline
C &X-ray & 200 & 3.3 $\times 10^{12}M_{\odot}$\\
& center & & \\
\hline
D &lensing & 350 & 5.9 $\times 10^{12}M_{\odot}$\\
& center & & \\
\hline
\end{tabular}
}
\end{table}

Figure \ref{change-p} shows the convergence map changing as p is varied.
Here, we choose $r_h=50kpc$ for the vector fields.
For $p=0$, the solution is the same as MOND
There are two lensing centers that appear in the map,
but they are located approximately halfway between A and B,
and halfway between C and D. Thus, to reproduce the lensing
map of the bullet cluster would require vector-concentrations that are
very far outside the baryonic mass distributions.

However, as p is increased, the centers of the convergence contours move away from the
baryonic centers, and $\kappa$ is much larger than in MOND. This means that for $p$ near 1,
the vector fields play a role very much like dark matter, and are likely to
augment the lensing in very much the same way as do dark matter halos.

Figure \ref{change-rh} shows the convergence map changing as $r_h$ for the vector fields is varied.
To emphasize the effect of the vector field, we choose $p=0.99$.
Moving from panel a (top left) through panel d (bottom right),
$r_h$ changes from $200kpc$ to $10kpc$. For large $r_h$,
$\kappa$ is less than in MOND.
This is because (from equation \ref{eqn:xGEA})
the effective $a_0$ is
\be
a_{eff} = a_0\sqrt{1-p}.
\ee
The MOND acceleration regime applies only when $a< a_{eff}$.
If the scale height of the vector field ($r_h$) is very large,
then $a< a_{eff}$ only at very large distances;
at smaller distances, the gravity becomes Newtonian.
$\kappa$ is then much less than it would have been
under MOND. As $r_h$ is reduced,
a gravitational center appears and moves to the centers of the
vector fields and the $\kappa$ also becomes bigger.
Clearly, Figure \ref{change-p} and \ref{change-rh}
suggest that the space components of vector fields can replace dark matter.

\section{conclusion}
In this work, we analyzed the effects of gravitational lensing
caused by the vector field in the Generalized Einstein-Aether theory.
The results show that the vector field could cause the dissociation
of mass center and gravity center observed in the bullet cluster.
This would not be the vector field sourced by the matter,
but rather the vector field that seeded the growth of the matter perturbation.
We have not followed the evolution of this primordial
seed, but rather established the approximate properties
that it would have to explain the observations.

The problem with the vector field is that a particular
configuration of the vector field distribution is needed.
This is the same problem as the need for a particular
mass distribution of the dark matter. Future work will address
comparing the full weak lensing data instead of the convergence map only.
We note that, since TeVeS also includes the same type of vector field,
it is reasonable to believe that the gravitational acceleration of TeVeS
could be dissociated from mass centers by the vector fields.

The convergence map alone is not enough to distinguish this type of
vector field from dark matter. We need a more detailed analysis
to distinguish the two models,
eg. by looking at the lensing shear we may yet find that modified gravity
is distinguishable from dark matter.
Finally we note that this independent concentration of vector is
very much like a dark matter halo -- but a halo of vector
field rather than of particles. Thus MOND started as an effort
to build a theory in which what you see (baryons) is what you get
(i.e. sources all gravity). One way or another, even in modified
gravity theories, that seems not to be true.

\acknowledgments
It is a pleasure to thank Constantinos Skordis, Irit Maor, Pedro Ferreira and Thomas Zlosnik
for very useful discussions.
We thank the authors of \cite{bullet} for permission to reproduce
figure \ref{fig:bulletcluster}.
This work has been supported by grants from the US DOE, NASA and HEPCOS group at SUNY.

\end{document}